# Observation of spin-charge conversion in chemical-vapor-deposition-grown single-layer graphene

Ryo Ohshima [1,*], Atsushi Sakai [1,*], Yuichiro Ando [1,2], Teruya Shinjo [2], Kenji Kawahara [3], Hiroki Ago [3], and Masashi Shiraishi [1,2]

1. Graduate School of Engineering Science, Osaka Univ., Toyonaka 560-8531, Japan.
2. Department of Electronic Science and Engineering, Kyoto Univ., Kyoto 615-8531 Japan.
3. Institute for Material Chemistry and Engineering, Kyushu Univ., Fukuoka 816-8508, Japan.

* These two authors equally contributed to this work.

**Abstract**

Conversion of pure spin current to charge current in single-layer graphene (SLG) is investigated by using spin pumping. Large-area SLG grown by chemical vapor deposition is used for the conversion. Efficient spin accumulation in SLG by spin pumping enables observing an electromotive force produced by the inverse spin Hall effect (ISHE) of SLG. The spin Hall angle of SLG is estimated to be $6.1 \times 10^{-7}$. The observed ISHE in SLG is ascribed to its non-negligible spin-orbit interaction in SLG.

Spintronics using molecular materials has been regarded as an attractive field in spintronics [1]. Among molecules, graphene is regarded as having the greater potential in consequence of electrical [2–4] and dynamical [5] transport of a pure spin current being experimentally demonstrated at room temperature (RT). Because spin transport enables fabrication of spin devices, modulation of the spin signals of propagating pure spin current in graphene, i.e., an operation of spin transistor using graphene, was also reported [6]. To date, spin transport in other molecules at RT has not been adequately confirmed because the generation of a pure spin current at RT has not been achieved. However, successful dynamical spin injection in a conductive polymer, poly(3,4-ethylenedioxythiophene):poly(4-styrenesulphonate) (PEDOT:PSS), should be noted [7]. The authors confirmed the dynamical spin injection into PEDOT:PSS by observing its charge current generated by the spin-charge conversion, where the conversion is attributed to the inverse spin Hall effect (ISHE). As PEDOT:PSS consists of mainly light elements, such as carbon and hydrogen, the spin-orbit interaction (SOI) is expected to be comparatively small. However, the non-zero SOI and the sufficiently large accumulation of spin at the interface between PEDOT:PSS and yittrium-iron-garnet ($Y_3Fe_5O_{12}$, YIG), a spin source, enables spin-charge conversion via the SOI. This has greatly impacted molecular spintronics because it challenges the conventional understanding of spin physics in molecules. Concerning graphene, the SOI of defect-free and flat single-layer graphene (SLG) is almost negligible (1-50 μeV [8,9]). However, SLG on substrates loses its flatness and possesses ripples, resulting in roughly a 20-folds enhancement of the SOI [9]. Additionally, impurities in graphene can also provide an order in magnitude enhancement of the SOI [10]. The introduction of ripples and defects into graphene is almost inevitable at the current stage, and thus, one can expect spin-charge conversion in graphene to occur, if an efficient spin accumulation develops. Here, in this study, we shed light on the spin-charge conversion in graphene, which endows additional functionality in graphene-based spintronics because spin conversion in addition to spin injection and spin transport is now realized.

In our study, large-area SLG was grown by ambient pressure chemical vapor deposition (CVD) and transferred onto a single crystal YIG substrate ($1.5 \times 3$ mm$^2$). Details of the CVD and transfer processes are described in ref. 11 and supplementary material [12]. Dynamical spin pumping was used to inject spin into the SLG; here YIG is the source of spins [13]. The principle of dynamical spin injection is as follows: Magnetization dynamics in a ferromagnet are described by the Landau-Lifshitz-Gilbert (LLG) equation,

$$\frac{dM}{dt} = \gamma H_{eff} \times M + \alpha \frac{M}{M_s} \times \frac{dM}{dt}, \qquad (1)$$

where $\gamma$, $M$, $H_{eff}$, $\alpha$, and $M_s$ are the gyromagnetic ratio of the ferromagnet, the time-dependent magnetization of the ferromagnet, an external magnetic field, the Gilbert damping constant, and saturation magnetization of the ferromagnet, respectively. The first and the second terms are the field term that describes magnetization precession and the damping term that describes damping torque, respectively. Ferromagnetic resonance (FMR) occurs when microwaves ($f$=9.6 GHz in this study) are applied to the ferromagnet. The damping torque is suppressed in FMR, which induces the pumping of spins into the SLG in consequence of spin angular momentum conservation. The pumped spins accumulate in the SLG. The YIG substrate is a ferromagnetic insulator with a band-gap of 2.7 eV, which helps to mitigate unnecessary electromotive forces from a conductive ferromagnet, such as NiFe [14], in spin pumping experiments. The YIG substrate, grown by liquid phase epitaxy, is commercially available (GRANOPT, Akita, Japan). Its surface roughness estimated to be 2.7 nm by atomic force microscopy impedes efficient spin pumping from the YIG substrate. Indeed, no electromotive force was observed from Pt on the as-grown YIG during the dynamical spin pumping, although Pt has a large SOI. To solve this problem, we polished the YIG. A gliding process using a suspension of alumina particles (0.05 μm in diameter) reduced the roughness down to 1.2 nm [12]. The results of polishing were experimentally checked by measuring the electromotive force from a 10-nm-thick Pt (Fig. 1(a)). The electromotive force was observed from the Pt. The electromotive force as a function of the angle of the external magnetic field exhibits a characteristic feature, i.e., its behavior correlates with the symmetry of the ISHE ($J_s \sim J_c \times \sigma$). This symmetry in the electromotive force is direct evidence of successful spin pumping indicating that surface conditions of the YIG substrate were improved by polishing. Indeed, the spin mixing conductance in this experiment was estimated to be $2.7 \times 10^{18}$ m$^{-2}$, which is better than the previous reported values [13]. The microwave power dependence of the electromotive forces from the Pt is shown in Fig. 1(b), and the electromotive force is linearly dependent on power up to ca. 1 mW. The linear dependence also supports spin pumping from the YIG into the Pt. We used the polished YIG to transfer the SLG onto the YIG. Here, the importance is the polarity of the ISHE signals from the Pt, i.e., an electron is a spin carrier in Pt. If the spin carrier is changed to holes, the polarity of the ISHE should be reversed, which holds in SLG (discussed below). Hence, this result provides a sufficient material to discuss and corroborate the spin-charge conversion in SLG.

Figure 2(a) shows a schematic of a spin-to-charge conversion device using the SLG. For detection of the electromotive force from the SLG, two 1-mm-wide Au electrodes were attached, set 1 mm apart. The *I-V* curve for the SLG is shown in Fig. 2(b). A linear dependence of current with voltage was observed, indicating that the SLG is conductive and an Ohmic contact between the Au and the SLG was established. The Hall measurement indicates that the carriers in the SLG are holes (see Fig. 2(c)); i.e., the SLG is a p-type zero-gap semiconductor. Indeed, SLGs fabricated using the same method exhibited p-type characteristics under a zero gate voltage in transport measurements [11,15] consistent with our Hall measurement result. We deduce that the p-type character is due to (1) water or oxygen adsorption or both, and/or (2) a doping effect due to the work function difference between YIG and SLG as in the SLG/SrTiO$_3$ [16].

Since spurious signals due to thermal effects by the microwave irradiation (not due to the ISHE) were superimposed to the detected electromotive forces, we introduced the following procedure in order to eliminate the thermal signals from the detected electromotive forces: the point is that the thermal effects exhibit electromotive forces with the same polarity with the external magnetic field of $\theta$ and $\theta$+180°, whereas the ISHE exhibits ones with the opposite polarity because of its symmetry ($J_s \sim J_c \times \sigma$). Now, we set the detected signal including the thermal signals at the external magnetic field of $\theta$ to be $V(\theta)$. The electromotive force due to the ISHE at $\theta$ can be obtained by using $\{V(\theta)-V(\theta$ +180°)\}/2 because of the above mentioned reason. We believe that the Nernst effect and the spin Seebeck effect in SLG are almost negligible, because SLG consists of one atomic layer and thermal gradient perpendicular to the SLG plane should be quite small, and also because the thermal conductivity of SLG is very high. The results are shown in Fig. 3(a). Although electrons were the spin carriers in Pt as discussed above, holes are the spin carriers in the SLG. Thus, the sign of the electromotive forces from the SLG at 0° was positive, the opposite to that observed in the Pt. Because saturation of the FMR signals of the YIG easily takes place under microwave irradiation, the maximum power that can be applied was 1 mW. However, the power dependence of the electromotive forces from the SLG exhibits linear dependence below 1 mW (Fig. 3(b)). This supports the notion that the observed electromotive force from the SLG can be ascribed to the ISHE. Furthermore, its angular dependence also exhibits the characteristic feature of the ISHE, as shown in Fig. 3(c). The external magnetic filed was rotated from 0° to 360° for this experiment. As explained above, the electromotive force due to the ISHE at $\theta$ was obtained by using $\{V(\theta)-V(\theta+180°)\}/2$, and for example, the data at $\theta$=30° is obtained by using $\{V(30°)-V(210°)\}/2$. The electromotive forces due to the ISHE apparently exhibit $\cos\theta$ dependence. All of the observations verify the successful conversion from the injected pure spin current to a charge current in the SLG, which we attribute to the non-negligible SOI in the SLG.

This conversion enables the spin Hall angle of the SLG to be calculate based on a model proposed in a previous study by Ando et al. [7]. To implement a precise quantitative analysis, we calculated the spin Hall angle using the values of the irradiation power under 0.4 mW. The electromotive force from the Pt on the YIG can be expressed as

$$V_{ISHE} = \theta_{SHE}^{Pt} \frac{\hbar\omega}{4e}\left(\frac{w}{d_N}\right)\frac{\tanh(d_N/\lambda_N)\tanh(d_N/2\lambda_N)}{1+\Gamma^2\tanh^2(d_N/\lambda_N)}\left(\frac{\gamma h_{ac}}{\alpha\omega}\right)^2, \qquad (2)$$

where $\theta_{SHE}^{Pt}$ is the spin Hall angle of Pt (0.04), $\hbar$ the Dirac constant, $\omega$ (=2$\pi f$) the angular frequency of the microwave, $e$ the charge, $w$ the width of the YIG substrate (3 mm) as shown in Fig. 2(a), $d_N$ the thickness of the Pt (10 nm), $\lambda_N$ the spin diffusion length of Pt (7 nm), $\gamma$ the gyromagnetic ratio ($1.78 \times 10^7$ G$^{-1}$s$^{-1}$), $h_{ac}$ the amplitude of the microwave field (0.028 G under the microwave power of 0.4 mW), and $\alpha$ the Gilbert damping constant of YIG ($6.7 \times 10^{-5}$) [7,17]. The factor $\Gamma$ is equal to $(\hbar/SJ_{sd}\tau_s)(\lambda_N/a_{eff})$, where $S$ (=16) is an effective block spin per unit cell in the YIG, $J_{sd}$ the strength of the spin-exchange coupling between the magnetization and carrier spins at the interface, $\tau_s$ the spin lifetime of Pt (0.3 ps), $a_{eff} = v_e/a_s^2$ is the effective interaction range ($v_e$ (($1.23 \times 10^{23}$ cm$^{-3}$)$^{-1}$) is the volume per carrier, and $a_s$ (=1.24 nm) is the lattice constant for the localized spins at the interface) [7,13]. As the measured electromotive force from the Pt was 0.87 μV under microwave power of 0.4 mW, the value of $\Gamma$ is estimated to be 33.3. To separate the contribution from spin accumulation, we rewrite Eq. (2) in the form,

$$V_{ISHE} = \theta_{SHE}^{Pt}\left(\frac{w}{d_N}\right)\left(\frac{1}{e}\right)\tanh\left(d_N/\lambda_N\right)\tanh\left(d_N/2\lambda_N\right)\delta\mu_0,$$

$$\delta\mu_0 = \frac{\hbar}{4}\frac{\tanh\left(d_N/\lambda_N\right)}{1+\Gamma^2\tanh^2\left(d_N/\lambda_N\right)}[M \times \frac{\partial M}{\partial t}]_z \quad , \quad (3)$$

where $\delta\mu_0$ denotes the spin accumulation at the interface between the Pt (or the SLG) and the YIG substrate, $[M \times \frac{\partial M}{\partial t}]_z$ is the magnitude of the pumped spin angular momentum, and $M$ is the time-dependent magnetization. Hence, the $\delta\mu_0$ for the Pt was estimated to be

$$\delta\mu_0 \approx (\hbar/4)(1/880)[M \times \frac{\partial M}{\partial t}]_z.$$

To estimate $\delta\mu_0$ for SLG, we make the following assumptions. According to the *vertical* spin injection study using graphite, the ratio between in-plane and vertical conductivities of graphite was ca. $3.75 \times 10^3$ [18]. With an Elliot-Yafet-type spin relaxation mechanism governing the spin relaxation in SLG [19], the spin lifetime and diffusion length is proportional to the conductivity. The typical in-plain spin lifetime and diffusion length in conventional SLG are ca. 100 ps and 2 μm [2], although much longer spin coherence has been reported by using BN for obtaining flatness and high mobility of SLG [20] or bilayer graphene [21]. In this study, the SLG is placed on YIG, not on BN, and we assume the above-mentioned spin lifetime and diffusion length can be used. Consequently, spin lifetime and diffusion length for *vertical* spin transport are suppressed by factor $3.75 \times 10^3$, i.e., 26 fs and 0.53 nm. It should be noted that according to Ref. [18], electrons at 1.8 eV above the Femi level exhibit the spin transport across a 17 nm-thick graphite and the spin coherence for the electrons above the Fermi level was estimated to be 100 nm. In this study, electron spins are at the Fermi level, which is completely different in the experimental feature of Ref. [18]. Thus, we believe that our estimation of the spin transport parameters perpendicular to the SLG plane is plausible.

Now, $\Gamma J_{sd} \sim 0$ for SLG even though $\nu_e$ is almost at the lower limit for SLG ($1 \times 10^{12}$ cm$^{-2}$), and so, $\delta\mu_0$ of the SLG is estimated to be $\delta\mu_0 \approx (\hbar/4)[M \times \frac{\partial M}{\partial t}]_z$. The strong enhancement in spin accumulation enables the electromotive forces caused by the non-zero SOI in SLG to be detectable. With the measured electromotive force from the SLG being 0.11 μV under microwave power of 0.4 mW and the thickness of SLG being 0.335 nm, the spin Hall angle for SLG was estimated to be $6.1 \times 10^{-7}$, which is comparable to that of PEDOT:PSS.

Finally, we discuss why the electromotive force from the ISHE of SLG was successfully detected in this experimental set-up. In our previous study, the electromotive force from the SLG with NiFe was not detected despite similar mixing conductances ($10^{18}$–$10^{19}$ m$^{-2}$ in both studies). The electromotive force from the SLG with NiFe is expressed as $V_{ISHE} = \frac{w\theta_{SHE}^{SLG}\lambda_{SLG}\tanh(d_{SLG}/\lambda_{SLG})}{d_{SLG}\sigma_{SLG} + d_{NiFe}\sigma_{NiFe}}$, and the denominator includes the conductivity for NiFe, which is about one order of magnitude greater than that of SLG, and the thickness of NiFe was more than ten times larger than that of the SLG [5]. With regard to the SLG/YIG system the denominator becomes much smaller because YIG is an insulator. This allows a dramatic increase in the electromotive force for the SLG/YIG (by a factor of about 140). The electromotive force of several hundred nV was detected in our study, and we deduce that a quite small electromotive force, of order 1 nV, from the SLG was generated in the NiFe/SLG, which was below the detection limit. Hence, introducing YIG is crucial in detecting the ISHE in SLG.

In summary, we experimentally demonstrated the detection of the ISHE for SLG. The spin Hall angle of the SLG was estimated to be 6.1 × $10^{-7}$. This finding extends the functionality of graphene-based spin devices.

A part of this study was supported by Japan Science and Technology Agency (JST), Ministry of Education, Culture, Sports Science and Technology (MEXT) and Japan Society for the Promotion of Science (JSPS).

**Figure captions**

**Figure 1**

(a) Electromotive forces from Pt (10 nm) on a YIG substrate. A direction of an applied external magnetic field is indicated in the figure. The horizontal axis represents the difference between the applied magnetic field and the static magnetic field under FMR. The electromotive forces were observed at $\theta=0°$ and 180°, whereas no signal was observed at $\theta =90°$. (b) Microwave power dependence of the electromotive forces from the Pt. Linear dependence of the electromotive forces was obtained up to 1 mW. The dashed line represents a linear fit.

**Figure 2**

(a) Schematic of a sample structure of a spin-charge conversion device using SLG. The SLG was transferred onto a YIG/Gadolinium-Ion-Garnet (GGG) substrate, and two Au electrodes were attached. The Au electrodes were set 1 mm apart. (b) An *I-V* curve of the device. The linear dependence indicates the formation of an Ohmic contact between SLG and Au. Because YIG is an insulator, the conduction of the electric current is attributed to the conductive SLG. (c) Hall voltage from the SLG as a function of the magnetic field. The positive slope of the voltages apparently shows that the carrier in the SLG is a hole. The dashed line represents a linear fit.

**Figure 3**

(a) The FMR spectrum of the YIG at $\theta=0°$ (the upper panel). The static magnetic field was ca. 2605 G. The irradiation power was set to 0.8 mW. The electromotive force from the SLG on a

YIG substrate at $\theta=0°$ is shown in the bottom panels. The abscissa shows the difference between the applied and static magnetic fields producing FMR. As can be seen, the electromotive forces appear under the FMR of the YIG. The sign of the electromotive force at $\theta=0°$ is positive, as the charge carriers in the SLG are holes. (b) Microwave power dependence of the electromotive forces from the SLG. Linear dependence of the electromotive forces was obtained up to 1 mW. The dashed line is a linear fit. (c) Angular dependence of the electromotive forces from the SLG under 1 mW as a function of external static magnetic field. The dashed line follows a $\cos\theta$ dependence, which reproduces the experimental results.

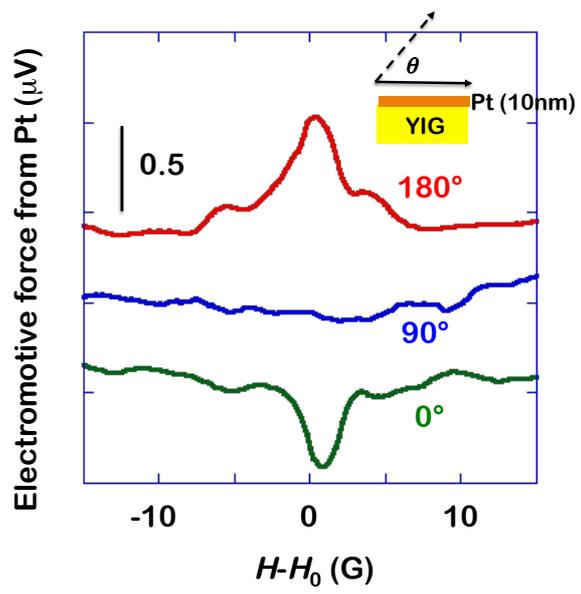

Fig. 1(a) R. Ohshima et al.

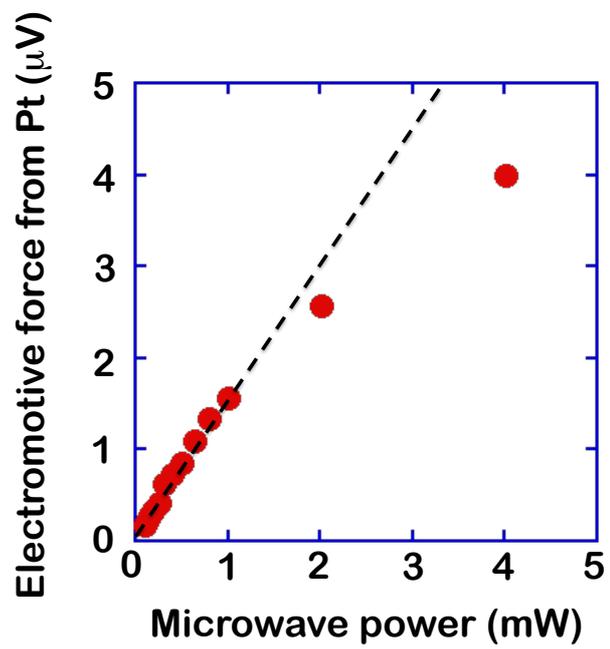

Fig. 1(b) R. Ohshima et al.

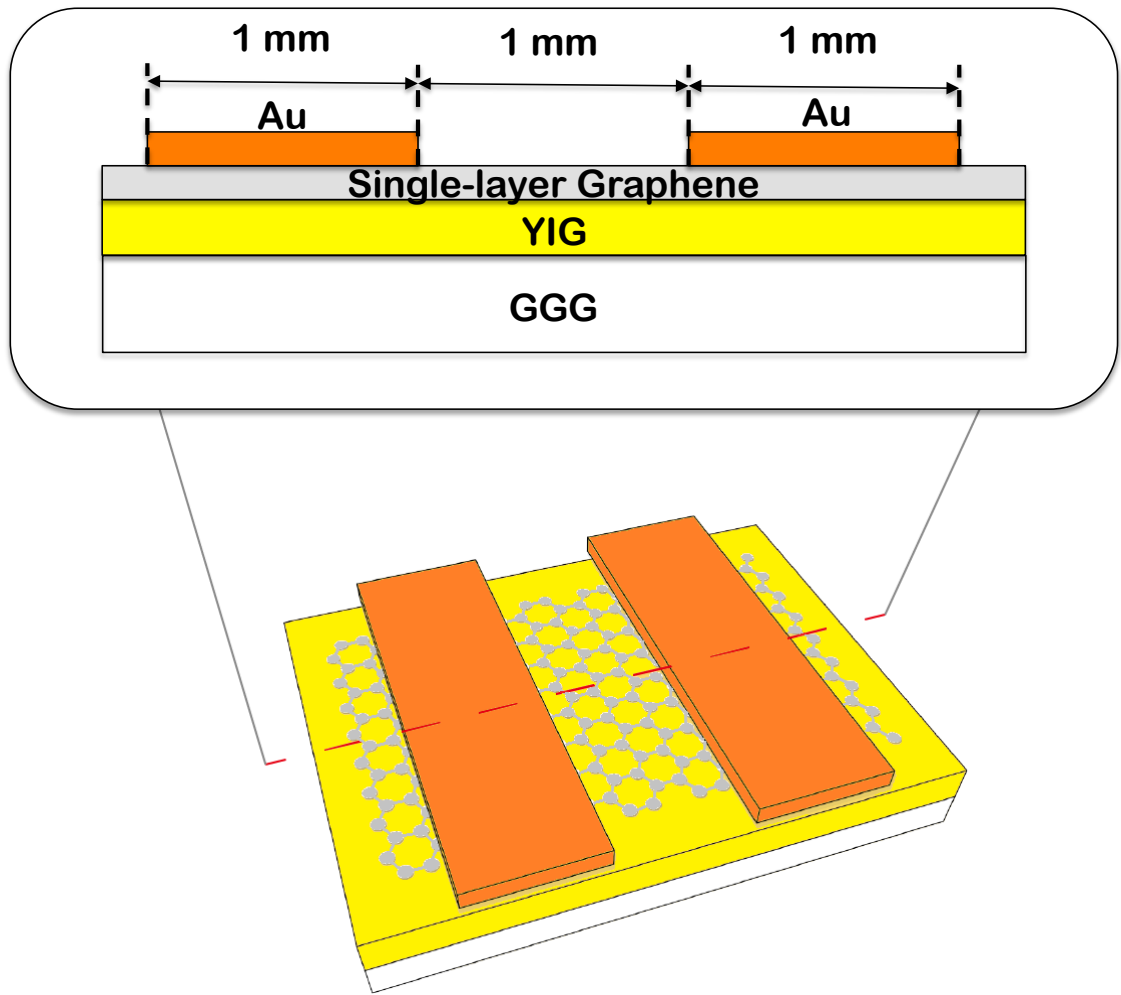

Fig. 2(a) R. Ohshima et al.

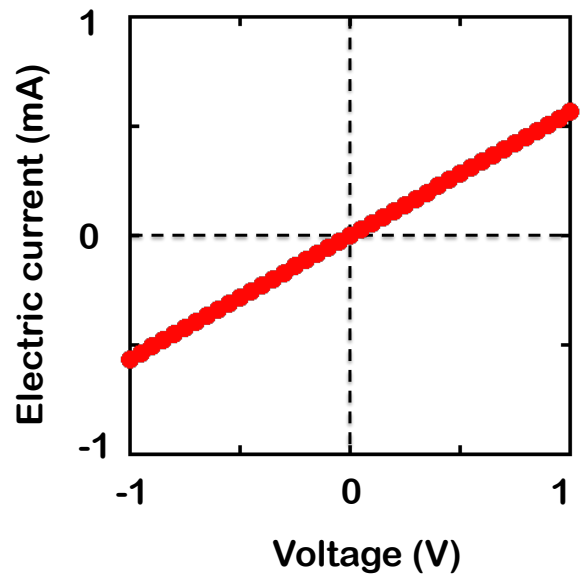

Fig. 2b R. Ohshima

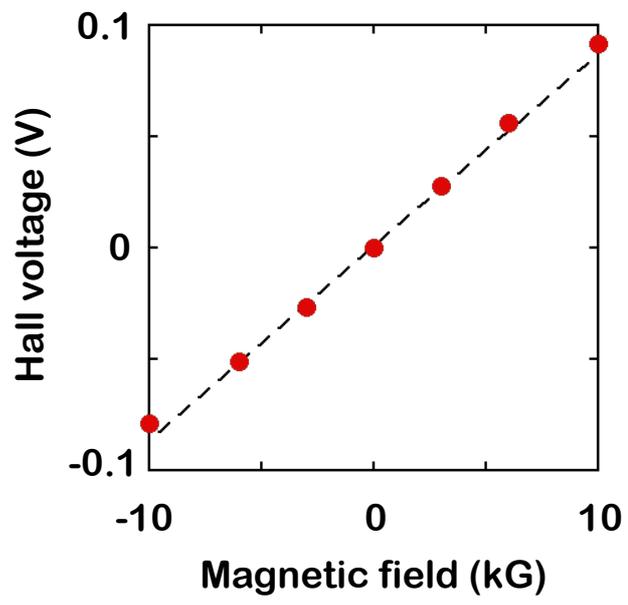

Fig. 2(c) R. Ohshima

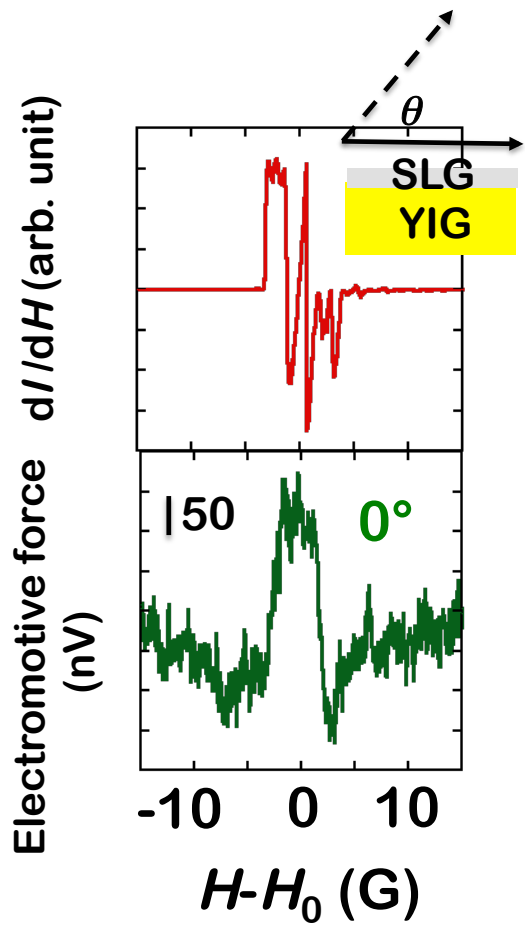

Fig. 3(a) R. Ohshima et al.

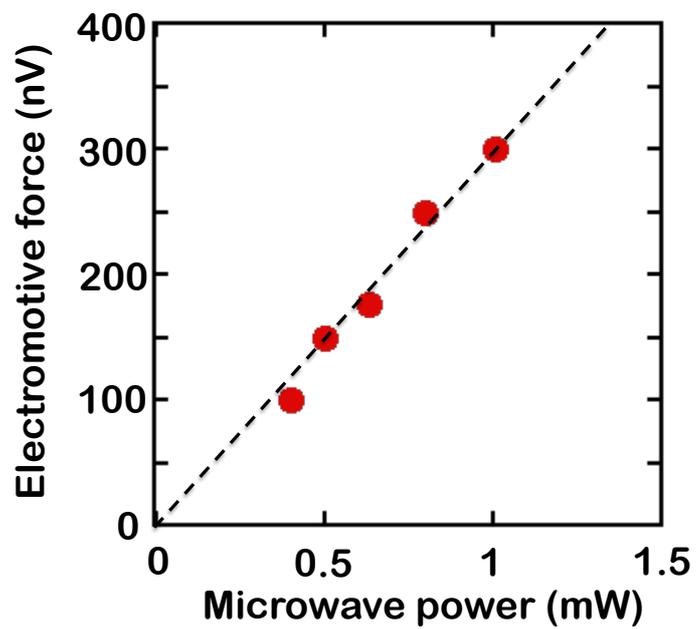

Fig. 3(b) R. Ohshima et al.

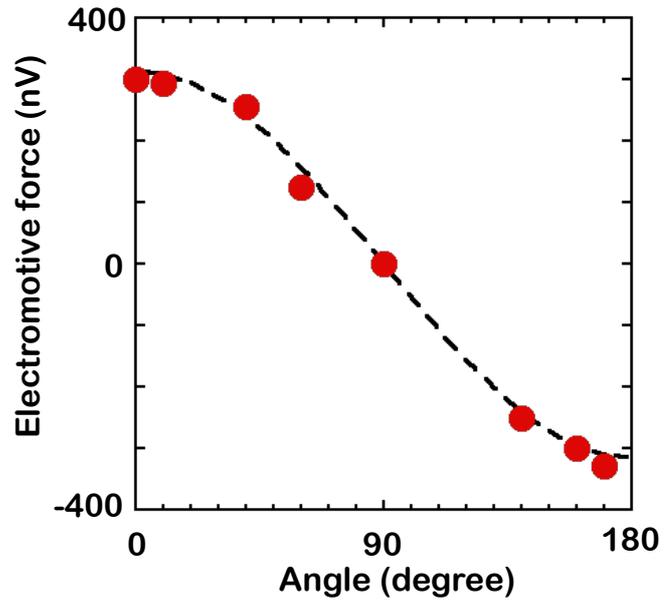

Fig. 3(c) R. Ohshima et al.